# Electron mass enhancement and magnetic phase separation near the Mott transition in double layer ruthenates

Jin Peng[1*], X. M. Gu[2], G. T. Zhou[2], W. Wang[2], J. Y. Liu[3], Yu Wang[3], Z. Q. Mao[3], X.S. Wu[2], and Shuai Dong[1]

[1]School of Physics, Southeast University, Nanjing 211189, China

[2]Collaborative Innovation Center of Advanced Microstrucutres, Lab of Solid State Microstructures, School of Physics, Nanjing University, Nanjing 210093, China

[3]Department of Physics and Engineering Physics, Tulane University, New Orleans, LA 70118, USA

*Corresponding author: e-mail jpeng@seu.edu.cn

We present a detailed investigation of the specific heat in $Ca_3(Ru_{1-x}M_x)_2O_7$ (M = Ti, Fe, Mn) single crystals. With different dopants and doping levels, three distinct regions are present, including a quasi-2D metallic state with an antiferromagnetic (AFM) order formed by ferromagnetic bilayers (AFM-b), a Mott insulating state with G-type AFM order (G-AFM) and a localized state with a mixed AFM-b and G-AFM phase. Our specific heat data provide deep insights into the Mott transitions induced by Ti and Mn dopings. We observed not only an anomalous large mass enhancement but also an additional term in the specific heat i.e. $C \propto T^2$ in the localized region. The $C \propto T^2$ term is most likely due to the long-wavelength excitations with both FM and AFM components. A decrease of Debye temperature is observed in the G-type AFM region, indicating a lattice softening associated with the Mott transition.

I. INTRODUCTION

Insulator-metal transition (IMT) accompanied by a huge resistivity change is one of the most studied phenomena in condensed matter physics area [1,2]. It not only has promising application in new generation information technology, but also is fundamentally important for studying the underlying physics of correlated electronic systems[3,4]. Interestingly, while there are many observations of insulator-metal transitions upon band filling or tuning of bandwidth $W$, there are few reports on Mott transitions driven by doping to a metal. One example of such impurity induced Mott transition is found in the prototype system of the Mott transition of $V_2O_3$ [5]. As $V^{3+}$ is replaced by $Ti^{3+}$, the system becomes metallic. However, the replacement of $V^{3+}$ by $Cr^{3+}$ instead enhances the Mott insulating phase [6]. Castellani *et al*., proposed that each $Cr^{3+}$ ions act as strong scattering center, increase of $Cr^{3+}$ content means increase of the nonpolar state weight versus the polar state, leading to a state of non-conductivity [7]. This type of transition should be a type of percolation problem. The above, and its related metal-insulator transition (MIT) has a wide technological potential. Later experimental efforts did support the view of phase separations near the Mott transition in this system. Another puzzling phenomenon observed in the system is the anomalous metallic state where the specific-heat coefficient $\gamma$ near the MIT show strong enhancement [8].

In this paper, we present another system which show an impurity-induced Mott insulating state, the Ruddelson-Popper-type layered ruthenates $(Sr, Ca)_{n+1}Ru_nO_{3n+1}$ [9]. This system displays various ground states including spin-triplet superconductivity ($Sr_2RuO_4$) [10,11], enhanced paramagnetic metallic state ($Sr_3Ru_2O_7$) [12], itinerant ferromagnetism ($SrRuO_3$) [13], antiferromagnetic (AFM) Mott insulating state($Ca_2RuO_4$) [14,15], quasi-two-dimensional metallic state with an AFM order ($Ca_3Ru_2O_7$) [16,17], and paramagnetic (PM) 'bad' metallic state ($CaRuO_3$) [18]. Our previous study shows that although the ground states are diverse, the doping effects of some $3d$ ions on Ru sites show consistency. That is, Mn and Ti dopants induce or enhance the Mott insulating state and AFM coupling, while Cr, Fe, or Co enhances the FM coupling [19]. For example, in double layer ruthenates, $Ca_3Ru_2O_7$, it shows an antiferromagnetic (AFM) transition at 56 K, which is then followed by a metal-insulator transition (MIT) at 48 K[20]. In plane resistivity recovers to metallic state below 30 K[16]. The AFM state below 56 K is characterized by ferromagnetic (FM) bilayers coupled antiferromagnetically along the $c$ axis. The spin direction switches from the $a$-axis (AFM-a) for $T_{MIT} < T < T_N$ to the $b$-axis (AFM-b) for $T < T_{MIT}$[17,21]. As few as 3% Ti doping or 4% Mn doping on the ruthenium site can tune the system from a quasi-2D metallic state with AFM-b order to a Mott insulating state with a G-type AFM order through a phase separation regions[19,22,23]. However, Fe doping on ruthenium site will not lead to a Mott insulating state, instead, the system show a localized electronic state with AFM-b and Incommensurate magnetic (ICM) structure coexistence[24]. Detailed phase diagram are shown in Figure 1. The schematic diagram of three magnetic structures AFM-b, G-AFM and ICM are shown in Fig. 4. Here, we report on the nature of this impurity-induced Mott transition revealed by specific heat measurements, including the phase separation and anomalous metallic state.

## II. EXPERIMENTAL DETAILS

Single-crystal samples were grown by floating zone methods. All samples used in our experiments were examined by X-ray diffraction (XRD) measurements and proven to be composed of pure bilayered phase. The successful doping of Ti, Mn, Fe ions into single crystals was confirmed by energy-dispersive x-ray spectroscopy (EDS). The real compositions are in general consistent with the nominal ones. Chemical formula in present are the real composition. Heat capacity measurements (2 K-200 K) were made in a physical property measurement system (Quantum Design) using the relaxation measurement. The masses of samples are measured by thermogravimetric analysis system with accuracy of 0.01 mg.

## III. RESULT AND DISCUSSION

The global temperature dependence of specific heat $C$ for Mn doped $Ca_3Ru_2O_7$ is shown in Figure 2a. The most notable high-$T$ feature s are the peaks at magnetic ordering temperature $T_N$ and MIT temperature $T_{MIT}$. For the parent compound $Ca_3Ru_2O_7$, it orders antiferromagnetically at $T_N$= 56 K, followed by an MIT and 48 K. A broaden "lambda anomaly" is observed at $T_N$; a sharp peak is observed at $T_{MIT}$. Indicating that the magnetic transition from paramagnetic to AFM-a is of second order and the MIT associate the magnetic transition from AFM-a to AFM-b is of first order. With Mn dopants induced, the peak corresponds to the MIT first move to e low temperature and split into two small peaks (1% Mn, 2% Mn), then gradually emerges as a bump (3% Mn, 4% Mn). At the mean time. The broaden "lambda anomaly" corresponding to the AFM ordering are barely affected by Mn dopants. When Mn doping level increase to above 5%, MIT and magnetic transition merges, leaving only one sharp peak at dramatically higher temperature. This is seen more clearly by subtracting a smooth background using a high-order polynomial function (Fig. 2b). Ti doped samples show similar behavior with Mn doped ones as shown in Fig. 2c, except that the critical concentration is 4% instead of 5%. For Fe doped samples, the "lambda anomaly" move to higher temperature, as well as the MIT move to low temperature more efficiently than Mn and Ti doped ones. $T_N$ reaches ~ 80 K for only 5% doping level. Besides, MIT and magnetic ordering never merges until the highest doping level synthesized successfully.

These observations in Fig. 2 clearly reveal three distinct composition groups for 3$d$ doped $Ca_3Ru_2O_7$ system. Group 1: The parent compound, which is characterized by a "lambda anomaly" and a sharp peak, separate the compound into high temperature paramagnetic (PM) phase, medium temperature AFM-a phase and low temperature AFM-b phase. Group 2: For low concentration, Ti or Mn doped, and all Fe doped samples. An intermediate magnetic phase emerges between AFM-a and AFM-b. This intermediate phase was characterized carefully by elastic neuter scattering measurements and proved to be a commensurate and incommensurate phase coexistence. Group 3: For high concentration Ti or Mn doped region, the single strong peak separate high temperature PM metallic phase and low temperature G-AFM Mott insulating state. We will further discuss these regions in details by analysis of the low temperature specific heat.

The low temperature regions (2K < $T$ < 10K) is shown in Fig. 3 for selected compositions. The data are plotted as $C/T$ vs $T^2$ and can be divided by contribution from different excitations as follows: $C = C_V + C_e + C_M + C_h$, where $C_V$ is the lattice contribution which is equal to $\beta T^3$, $\beta$ is given in Debay model by $\beta = 234 N k_B / \theta_D^3$, $\theta_D$ indicates Debay temperature. $C_e$ is the electron contribution which is equal to $\gamma T$, the sommerfeld coefficient $\gamma$ is given by $\gamma = \pi^2 k_B^2 N(E_F)/3$, here $N(E_F)$ means the density of states (DOS) at fermi level. $C_M$ is the contribution by magnetic fluctuations. $C_h$ is the contribution from ions' nucleus which is usually proportional to $T^{-2}$. Compositions in group 1 ($Ca_3Ru_2O_7$) and group 3 (10% Ti and 8% Mn) can be simply fitted by only the first two factors $C = \gamma T + \beta T^3$ as shown by the red curves in Fig. 3. However, compositions in group 2 (1% Mn, 2% Ti, 3% Fe, 5% Fe) show an obvious downward turn at low temperature limit, indicating a low order term than $\beta T^3$. The data of these compositions are thus fitted to

$C = \gamma T + BT^2 + \beta T^3$. Parameters $\gamma$ and $B$ are summarized in Fig. 4. Debay temperatures $\theta_D$ derived from parameter $\beta$ are plotted as a function of doping level in Fig. 5.

The $T^2$ contribution to specific heat have been observed in manganites La$_{1-x}$Sr$_x$MnO$_{3+\delta}$ when it is an A-type antiferromagnet [25]. Considering a model with linear dispersion for the planar ferromagnetic excitation and quadratic dispersion for the linear antiferromagnetic excitation, the combining dispersion relationship yields the low-temperature magnetic contribution to the specific heat, offering a possible candidate for $T^2$ term. However, this interpretation cannot be directly applied to our system. First, the proposed model is for three dimensional infinite layer perovskite structure (ABO$_3$ type). The studied system here is belong to quasi-2D double layer perovskite structure. AFM-b is ferromagnetic double layers coupled antiferromagnetically along $c$-axis. Magnetic excitation not only include planar ferromagnetic and linear antiferromagnetic, but also linear ferromagnetic. This will yields different dispersion relationship with A-type AFM structure. Second, the $T^2$ term was not observed in parent compound Ca$_3$Ru$_2$O$_7$ which also show AFM-b ground state.

We further found that some other compounds which do not have A-type AFM also have $T^2$ term, such as electron doped CaMnO$_3$ and La$_{1-x}$Sr$_x$CoO$_3$ [26,27]. They interpreted it as being due to the long-wavelength excitations with both FM and AFM components, due to the magnetic phase separation. This seems similar to our situation. In group two compositions, the ground states of Mn and Ti doped one exist the phase separation between AFM-b and G-type AFM. Fe doped samples consist both commensurate AFM-b phase and incommensurate phase formed of a cycloidal spiral spin structure. Regrettably, theoretic study in terms of the magnetic entities are still lacking.

Another puzzling feature is the strong enhancement of sommerfeld coefficient $\gamma$ in group 2 compositions. The in-plane transport behaviors of group 2 samples show insulating/semiconducting temperature dependence with low residual resistivity. We named this region as localized state, in contrast to the Mott insulating state of group 3 compositions. As state above, $\gamma$ are proportional to $n(\varepsilon_F)$. However, non-zero value of $\gamma$ in insulating materials were reported before. P. W. Anderson proposed that the linear specific heat component is a general feature on the glassy state, including spin glass, due to statistical distribution of localized "tunneling levels[28]. Interestingly, various studies of manganites reveal large $\gamma$ in insulating crystalline compositions. For example, in

Nd$_{0.67}$Sr$_{0.33}$MnO$_3$, a value of $\gamma = 25\ mJ/mol\ K^2$ was observed and in the electron-doped system La$_{2.3}$Ca$_{0.7}$Mn$_2$O$_7$, $\gamma = 41\ mJ/mol\ K^2$ .[29,30]. In the hole-doped LaMnO$_{3+\delta}$, $\gamma$ reaches as high as $23\ mJ/mol\ K^2$ [31]. In La$_{0.2}$Sr$_{0.8}$MnO$_3$, $\gamma = 5.6\ mJ/mol\ K^2$ [32] .Obviously that the finite linear specific heat coefficient not only appear in spin glass phase.

In undoped Ca$_3$Ru$_2$O$_7$, the small value of sommerfeld coefficient ($\gamma \sim 1.7$ mJ/ Ru mol T$^2$) arising from the non-nesting fermi surface pockets survived. In Ru-site doped Ca$_3$Ru$_2$O$_7$, the itinerant Ru $t_{2g}$ electrons may be localized due to the potential fluctuations arising from cation substitution and spin-dependent fluctuations due to local deviations from AFM-b magnetic order. If the doping concentration is fairly low, the localization length may be fairly large. Charge carriers can thus hop through a number of Ru ions, defines limited length of bilayer FM clusters. On the other hand, the electron levels although localized, are not largely spaced in energy, allowing for thermal excitations that contribute with a linear term to specific heat. From Fig. 4, we noticed that enhanced $\gamma$ appears simultaneously with the $T^2$ term. As state above, the $T^2$ term is due to the long-wavelength excitations with both FM and AFM components in the magnetic phase separation region. The proposed scenario above is consistent with this picture.

Above scenario explain the non-zero sommerfeld coefficient in insulating state. Why this coefficient is one order larger than pristine compound. Actually, enhanced $\gamma$ near Mott transition is ubiquitous, especially the ones with antiferromagnetism. For example, the heavy fermion compounds, the high-$T$c cuprates, the Mott-Hubbard systems V$_2$O$_3$ and Ni(Se$_{1-x}$S$_x$)$_2$ [8,33,34]. In V$_2$O$_3$ system, metallic state can be achieved by Ti doping, under pressure, or V deficiency upon tuning of the one-electron bandwidth $W$ or band filling by doping. In the previous route (tuning of the one-electron bandwidth W), the electronic effective mass $m^*$ diverges at the MIT. For the later route (band filling by doping), $m^*$ decreases at the transition. Our case obviously more close to the former route. The mechanism of mass-enhancement can be magnetic polarons, lattice polarons, Coulomb-interaction effects or Van hove singularities near the fermi level.

In addition, we found that sommerfeld coefficient $\gamma$ and $T^2$ contribution B enhances simultaneously in group 2 region as shown in Fig. 4. As state above, the enhancements of sommerfeld coefficient $\gamma$ indicate the existence of charge carriers that can tunnel through the potential barrier among local minima. The $T^2$ contribution B are interpreted as being due to the long-wavelength excitations with both FM and AFM components. Usually, the FM component are due to double-exchange interaction mediated via the Hunds rule coupling between itinerant electrons and localized moments. AFM state are due to the super-exchange interaction based on the Goodenough-Kanamori rule. Therefore, the coexistence of FM and AFM components indicate there are still some itinerant electrons in group 2 region, which also explain the non-zero sommerfeld coefficient $\gamma$.

If Mn or Ti doping level further increased to group 3 compositions. Both $T^2$ term and sommerfeld coefficient $\gamma$ becomes negligible. This is understandable since that the ground states of this group's samples are G-type AFM Mott insulating state. For G-type antiferromagnetism, the contribution of magnetic excitation to heat capacity are in form of $T^3$. This will result an overestimation of Debay temperatures $\theta_D$ derived from parameter $\beta$. The overall magnitude of $\theta_D$ (450 K – 650 K) is typical for perovskite oxides of this type and are in consistent with other RP type ruthenates. The most surprising feature is the obvious reduction of $\theta_D$ in the G-type AFM region, especially considering that $\theta_D$ have already been overestimated in this region. This reduction suggests a significant change in lattice dynamics cross the Mott transition. It is natural to correlate this with the known discontinuity in unit-cell parameters. For compositions in the third group, lattices are significantly flattened than compounds in group 1 and group 2 from XRD and neutron measurements. The reduction of $\theta_D$ is also observed in $La_{1-x}Sr_xMnO_3$ single crystals, where it was interpreted in terms of lattice softening induced by dynamical short-range Jahn-Teller distortions[35]. Such distortions could also play the key role in this system.

IV. CONCLUSION

In summary, we have performed a comprehensive study of heat capacity study in Ru-site doped Ca3Ru2O7 system, and found that all compositions can be divided into three groups. We observed two conventional contributions to the heat capacity at low $T$. A lattice contribution ($\propto T^3$), and an electronic contribution ($\propto T$) in addition to an unexpected $T^2$ contributionin in group 2 composition. The doping dependence of parameters of these terms was analyzed in detail, providing a significant amount of information to the impurity induced Mott transition. In particular, we found evidence for the percolation nature of the Mott transition, and a large electron mass enhancement due to strong electronelectron correlations in the localized state. Additionally, we also found lattice softening in the Mott insulating state. The doping dependence of the $T^2$ contribution, electronic contribution ($\propto T$) were shown to provide a detailed picture of the evolution of the phase-separated state with doping. These results not only clarify the systematics of the magnetic phase separation in this system, they also reiterate the capabilities of specific-heat measurements as a powerful probe of magnetic inhomogeneity.

**Acknowledgements**

We gratefully acknowledge financial support from the fundenmental Research Funds for the Central Universities China (No. 3207028403). Work at Nanjing University was supported by the National key R&D program of China (Grant No. 2017YFA0303202). The sample growth and high-temperature specific heat measurements at Tulane University were supported by the U.S. Department of Energy under EPSCOR Grant No. DE-SC0012432 with additional support from the Louisiana Board of Regents.

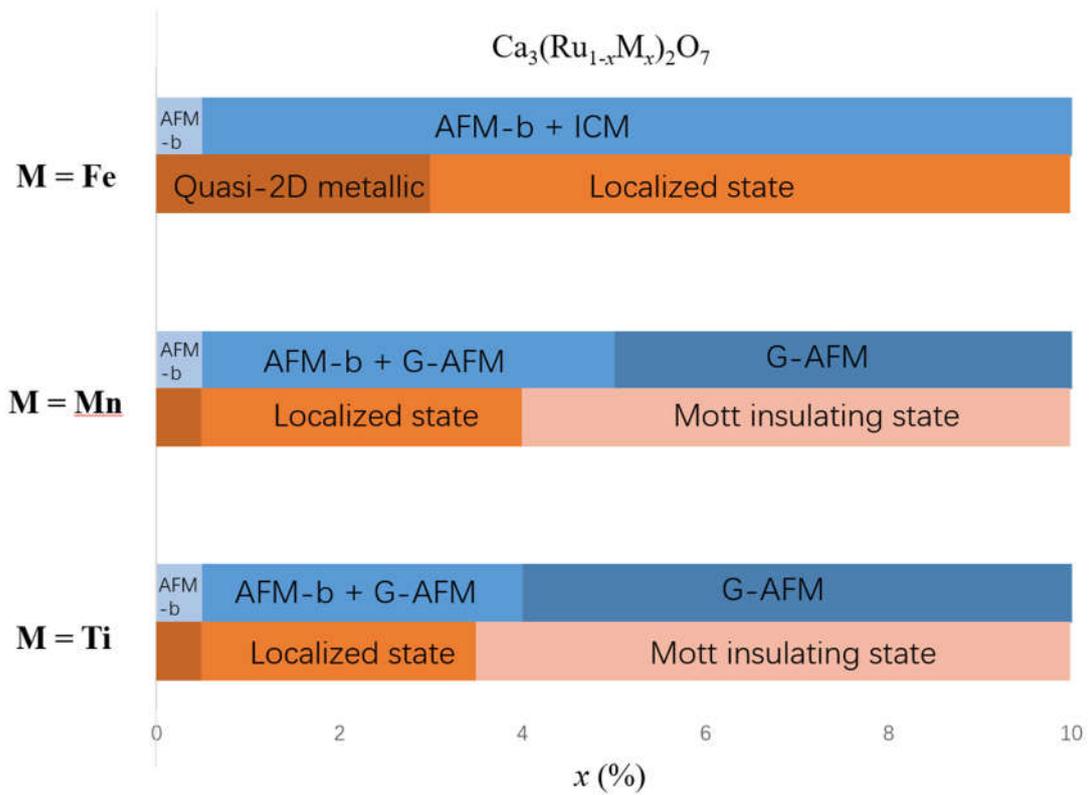

**FIG. 1.** Magnetic and electronic phase diagram of $Ca_3(Ru_{1-x}M_x)_2O_7$ (M = Ti, Fe, Mn), magnetic and electronic states are represented by different colored region and labeled.

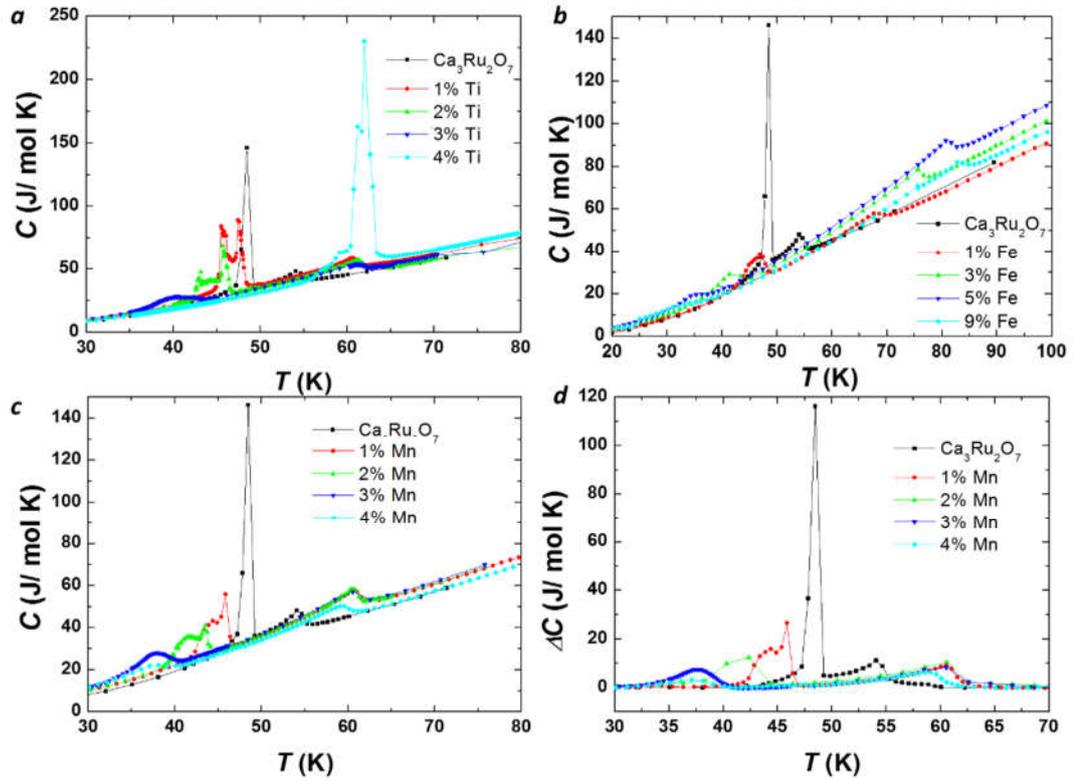

**FIG. 2**. Temperature dependence of the specific heat of: (a) pure, 1% Ti, 2% Ti, 3% Ti and 4% Ti doped $Ca_3Ru_2O_7$ (30 K – 80 K); (b) pure, 1% Fe, 3% Fe, 5% Fe and 9% Fe doped $Ca_3Ru_2O_7$ (30 K – 80 K); (c) pure, 1% Mn, 2% Mn, 3% Mn and 4% Mn doped $Ca_3Ru_2O_7$ (30 K – 80 K); (d) Excess (magnetic) specific heat extracted from the data in (c) by subtracting a smooth background as described in the text.

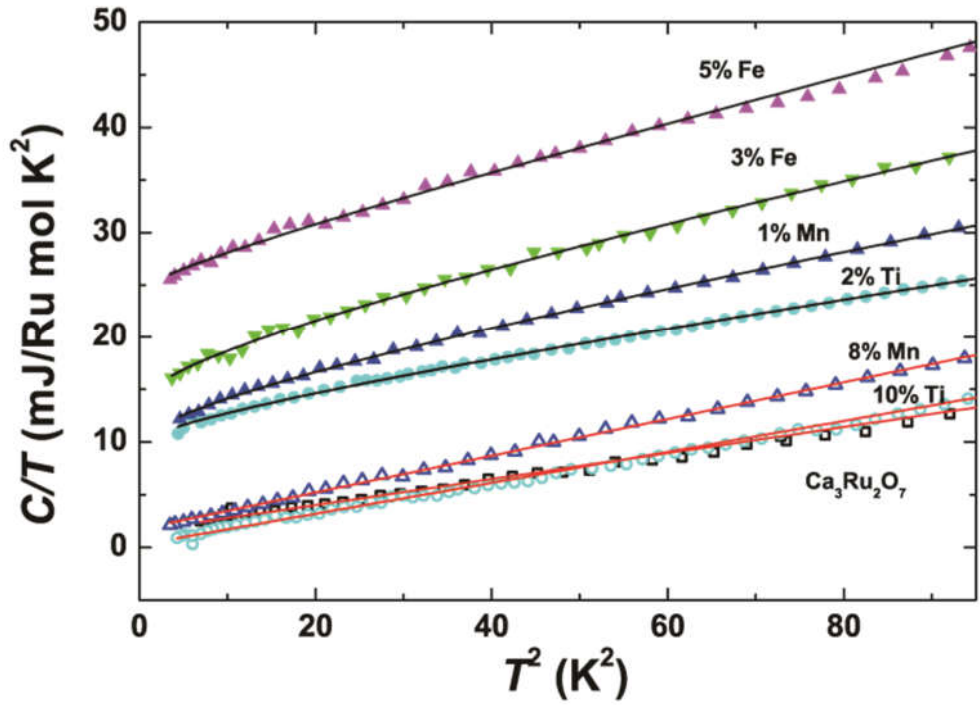

**FIG. 3**. Temperature dependence of the specific heat (2 K – 10 K) of 7 compositions (5% Fe, 3% Fe, 1% Mn, 2% Ti, 8% Mn, 10% Ti and Ca$_3$Ru$_2$O$_7$) plotted as $C/T$ vs $T^2$. The solid lines are fitted to $C = \gamma T + BT^2 + \beta T^3$, a model that is described in detail in text. The Adj. R-square parameters for these fittings are 0.98215, 0.99773, 0.99933, 0.99819, 0.99873, 0.99179, and 0.9876 respectively.

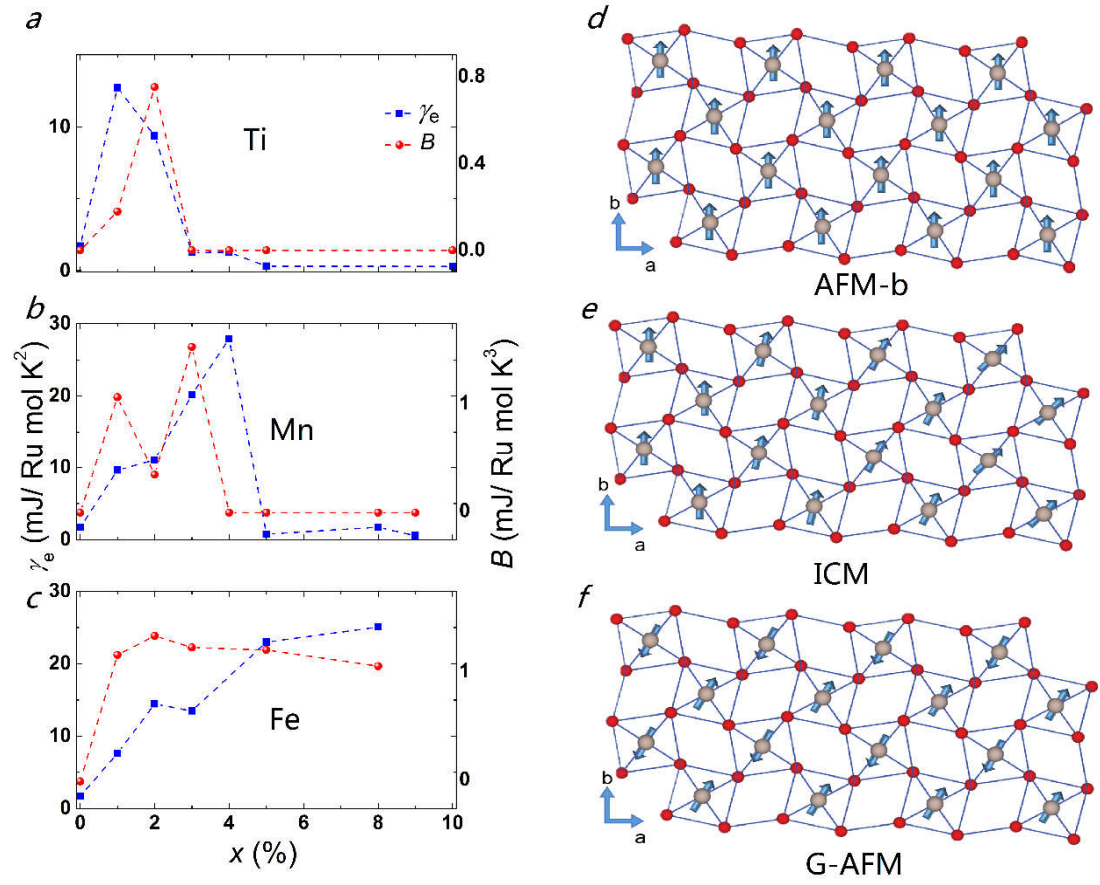

**FIG. 4**. Doping dependence of sommerfeld coefficient $\gamma$ and $T^2$ contribution $B$ to the specific heat for Ti (a), Mn (b) and Fe (c) doped $Ca_3Ru_2O_7$. (d) In-plane view of AFM-b magnetic strcuture. (e) In-plane view of Incommensurate magnetic structure. (f) In-plane view of G-AFM magnetic structure.

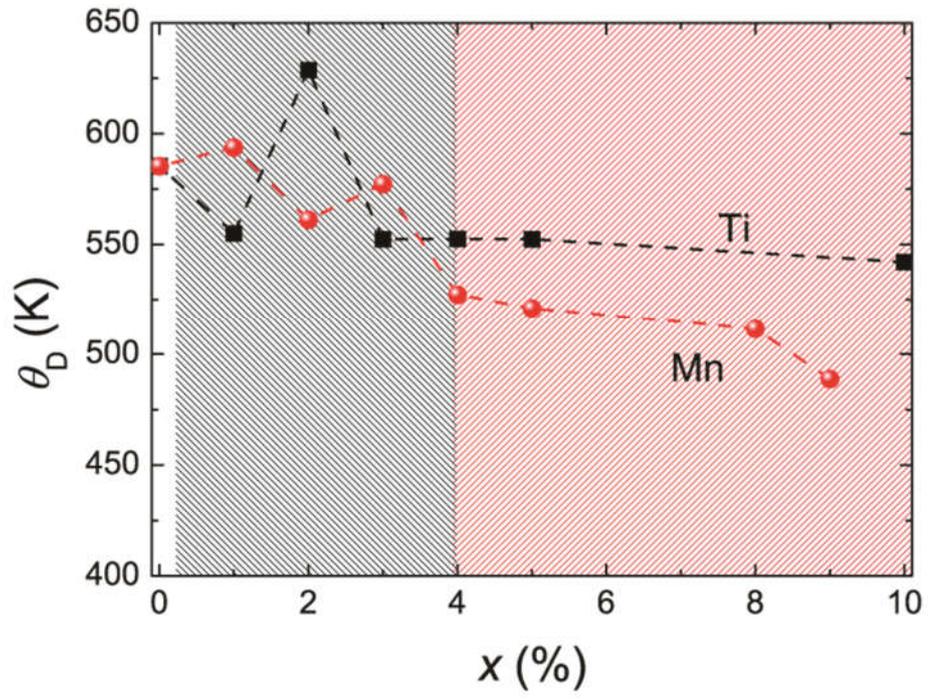

**FIG. 5**: Doping dependence of the Debye temperature $\theta_D$ for Mn and Ti doped $Ca_3Ru_2O_7$.